# SNR-adaptive OCT angiography enabled by statistical characterization of intensity and decorrelation with multi-variate time series model

Luzhe Huang, Yiming Fu, Ruixiang Chen, Shanshan Yang, Haixia Qiu, Xining Wu, Shiyong Zhao, Ying Gu, and Peng Li*

*Abstract*—In OCT angiography (OCTA), decorrelation computation has been widely used as a local motion index to identify dynamic flow from static tissues, but its dependence on SNR severely degrades the vascular visibility, particularly in low-SNR regions. To mathematically characterize the decorrelation-SNR dependence of OCT signals, we developed a multi-variate time series (MVTS) model. Based on the model, we derived a universal asymptotic linear relation of decorrelation to inverse SNR (iSNR), with the variance in static and noise regions determined by the average kernel size. Accordingly, with the population distribution of static and noise voxels being explicitly calculated in the iSNR and decorrelation (ID) space, a linear classifier is developed by removing static and noise voxels at all SNR, to generate a SNR-adaptive OCTA, termed as ID-OCTA. Then, flow phantom and human skin experiments were performed to validate the proposed ID-OCTA. Both qualitative and quantitative assessments demonstrated that ID-OCTA offers a superior visibility of blood vessels, particularly in the deep layer. Finally, implications of this work on both system design and hemodynamic quantification are further discussed.

## I. Introduction

UTILIZING the intrinsic motion of red blood cells (RBCs), optical coherence tomography angiography (OCTA), as a functional extension of OCT, enables a non-invasive, label-free 3D visualization of vasculature and perfusion down to capillary level by measuring the RBC-induced dynamic changes. The motion contrast provided by OCTA eliminates the requirement of exogenous contrast injection, allowing a fast and frequent vascular examination, and thus a broad spectrum of OCTA applications have been found in scientific research and clinics [1], such as ophthalmology [2]-[4], dermatology [5], [6], neurology [7] and oncology [8].

However, the measured motion index is not simply dependent on the actual motion magnitude of RBCs, but also influenced by the local backscattered light intensity from RBCs and noise level, i.e., local signal-to-noise-ratio (SNR). In general, the prevalent OCTA analyzes temporal changes in-between successive tomograms acquired at the same location as a motion index, and removes surrounding tissues according to the measured motion index. To accommodate the motion index measurement in various system configurations and situations, a variety of OCTA methods have been developed using different operations (including subtraction, variance, and decorrelation) of different signals (including intensity-based [9], [10], phase-based [11], [12], and complex-based [5], [13]-[20]). Obviously, the SNR-dependent motion index would result in significant misclassification of dynamic and static regions, severely degrading the vascular visibility and hindering the interpretation of imaging outcomes such as hemodynamic quantification [3], [4], [21]-[25].

Thus, it is necessary to combine both measured motion and intensity features to achieve precise classification of the dynamic flow from static tissues. The earliest and simplest approach is to directly eliminate the low-SNR regions from final angiograms. In the decorrelation-based OCTA, an intensity mask has been widely used to directly remove noise [18], [19], [26], but the empirical threshold either removes low-SNR flow by mistake or leaves residual static tissues inducing noisy background. As a sophisticated solution, modified decorrelation algorithms have been reported aiming to retrieve the true correlation of signal [27] or to correct the noise bias [20], whereas these modified algorithms involve complicated estimation of OCT parameters. Different from generating SNR-corrected OCTA signal, another direction is to generate a SNR-adaptive (or motion thresholding) classification. Due to the lack of a clear mathematical relation of OCTA to SNR, the numerical approaches (including fitting [28], learning [16], and regression [29]) were used for classification. A depth-adaptive motion classifier was achieved by fitting the OCTA statistical model to the measured depth-resolved histograms [28], while the statistical models used in their study is only suitable for

Manuscript received January 1, 2019; revised March 10, 2019 and April 08, 2019, accepted April 09, 2019. This work is supported by Zhejiang Provincial Natural Science Foundation of China (LR19F050002, LZ15F050002), National Natural Science Foundation of China (61475143, 61835015), Fundamental Research Funds for the Central Universities (2018FZA5003), National Key Research and Development Program of China (2017YFA0700501).

L. Huang, Y. Fu, R. Chen, S. Yang and P. Li are with the State Key Lab of Modern Optical Instrumentation, College of Optical Science and Engineering, Zhejiang University, Hangzhou, Zhejiang 310027, China. (*Corresponding author, e-mail: peng_li@zju.edu.cn).

H. Qiu and Y. Gu are with the Department of Laser Medicine, Chinese PLA General Hospital, Beijing 100853, China.

X. Wu and S. Zhao are with Tianjin Horimed Medical Technology Co., Ltd, Tianjing 300300, China.

subtraction-based OCTA (e.g. OMAG). A feature space OMAG has been proposed by Zhang et al. [16], in which a classification map is determined by machine learning of flow phantom data. However, the real blood flow has a broad variance in the OMAG signal and the differences between the flow phantom and real blood flow would inevitably affect the accuracy of the classification map. A reflectance-adjusted decorrelation threshold was proposed by Gao et al. [29], in which they calculated the decorrelation values (SSADA signals) at different reflectance levels in the foveal avascular zone (FAZ), and concluded a numerical relation between decorrelation and reflectance through linear regression. However, in addition to the reflectance (i.e. intensity), other possible factors such as the system noise and average kernel size have not been comprehensively considered in the acquired linear relation and the avascular region is hardly available in most biomedical imaging, resulting in a limited adaptability and generality to different systems and situations. Thus, a thorough mathematical characterization of SNR and OCTA signal might enable a straightforward approach to the problem of SNR dependence of OCTA signal.

The complex-decorrelation technique makes full use of both intensity and phase information of OCT signal, and thus provides superior motion contrast [19], [30]. Benefitting from its similarity calculation between two frames, the decorrelation-based angiography is intrinsically insensitive to disturbance caused by overall variation of the light source intensity [26]. In addition, the decorrelation value is less sensitive to the Doppler angle and thus has been used for hemodynamic quantification [21], [25].

In this work, we propose a novel, SNR-adaptive ID-OCTA enabled by the relation between inverse SNR (iSNR) and decorrelation (ID). In Section II, we introduce the mathematical fundament of ID-OCTA: developing the multi-variate time series (MVTS) model, deriving the asymptotic relation between ID features, analyzing the variance of the asymptotic distribution and developing a SNR-adaptive ID-OCTA method based on the statistical properties obtained above. Then, in Section III, flow phantom and human skin experiments were performed to show benefits of ID-OCTA both quantitatively and qualitatively. Finally, the value of our work, advices for practical implementation and further improvement of the proposed method are presented in Section IV.

## II. METHOD

In this section, we explicate the theoretical fundament of the proposed ID-OCTA. Firstly, the asymptotic ID relation is derived mathematically by a MVTS model. Secondly, the variance of the asymptotic distribution in static regions is analyzed numerically. Then the population distribution of the static voxels can be analytically determined in a 2D space with iSNR and decorrelation dimensions. Accordingly, a SNR-adaptive ID-OCTA method is developed in the third part by removing the static and noise voxels at all SNR.

### A. Inverse SNR-Decorrelation (ID) asymptotic relation

In the proposed ID-OCTA, local complex decorrelation is computed between successive B-frames taken at the same location with a 4D spatio-temporal average kernel, and used as a motion index to identify the RBC-induced dynamic changes:

$$D = 1 - \frac{C}{I}, \quad (1)$$

$$C = \frac{1}{(T-1)M} \sum_{m=1}^{M} \sum_{t=1}^{T-1} X(m_0 + m, t_0 + t) \cdot X^*(m_0 + m, t_0 + t + 1), \quad (2)$$

$$I = \frac{1}{TM} \sum_{m=1}^{M} \sum_{t=1}^{T} X(m_0 + m, t_0 + t) \cdot X^*(m_0 + m, t_0 + t). \quad (3)$$

Here $C$ is the local first-order sample auto-covariance at spatial index $m_0$ and $I$ is the local zeroth-order sample auto-covariance, or generally called intensity. $t_0$ is the temporal index and $*$ means the complex conjugate. For simplicity, the spatial index $m$ is used to denote $(z, x, y)$ and $M$ denotes the kernel size in three spatial dimensions. $T$ is the number of repetitions for B-scans at the same location. $X(m, t)$ is the complex OCT signal subjected to additive random noise [20]

$$X(m, t) = A(m, t) \cdot p(m, t) + n(m, t), \quad (4)$$

where $A(m, t)$ is the true OCT signal of the measured sample, $p(m, t)$ is an additional, slowly varying phasor with unit amplitude due to the instability of system and small motion of the sample, and $n(m, t)$ refers to the random noise.

A MVTS model was utilized here to mathematically derive the statistical properties of OCTA signal. In this model, we suppose that the noise $n$ and true signal $A$ have certain fundamental properties locally, which is essential for the following derivation. These assumptions generally hold for measured samples.

1) $n(m, t)$ are mutually independent white noises, and

$$\begin{aligned} \mathrm{E}(C) &= \mathrm{E}\left[\frac{1}{(T-1)M} \sum_{m=1}^{M} \sum_{t=0}^{T-1} X(m_0 + m, t_0 + t) \cdot X^*(m_0 + m, t_0 + t + 1)\right] \\ &= \frac{1}{(T-1)M} \sum_{m=1}^{M} \sum_{t=0}^{T-1} \mathrm{E}[A(m_0 + m, t_0 + t) \cdot A^*(m_0 + m, t_0 + t + 1) \\ &\quad + A(m_0 + m, t_0 + t) p(m_0 + m, t_0 + t) n^*(m_0 + m, t_0 + t + 1) \\ &\quad + A^*(m_0 + m, t_0 + t + 1) p^*(m_0 + m, t_0 + t + 1) n^*(m_0 + m, t_0 + t) \\ &\quad + n(m_0 + m, t_0 + t) n^*(m_0 + m, t_0 + t)] \end{aligned} \quad (5)$$

uncorrelated with the true signal. Here, we use E, Var to denote expectation and variance operators respectively. That means $E[n(m,t)] = 0$, $Var[n(m,t)] = E[n(m,t)n^*(m,t)] = s^2$, and $E[A(m,t)n^*(m',t')] = 0$, $\forall t, t', m, m'$, where $s^2$ refers to the variance of white noises.

2) The true OCT signals are mutually independent and identically distributed, stationary and ergodic multivariate time series. So $A(t|m), \forall m$, are time series with the same distribution, where $t|m$ means fixed $m$ and variable $t$. Moreover, we use $u, v^2$ to denote the mean value and variance of the time series respectively, that is, $E[A(m,t)] = u, Var[A(m,t)] = E[A(m,t)A^*(m,t)] = v^2, \forall m, t$.

3) Given $m$, the true OCT signal $A(t|m)$ is partially correlated with respect to $t$. Thus, we suppose its first-order covariance $E[(A(m,t)A^*(m,t+1)] = r_1, \forall t, m$. In addition, we denote its first-order correlation coefficient as $k = \frac{r_1}{v^2}$, and consequently $0 \leq |k| \leq 1$.

In fact, for static and pure noise voxels, the signal is supposed temporally invariant so $k = 1$. Besides, it is worth noting that even though one can further assume specific distributions about the series above, e.g., Rayleigh distribution for $A(m,t)$ and Gaussian distribution for noises, they are not necessities in our method, which broadens the applicability of this model to different apparatuses and samples.

With assumptions listed above, the expectation of first order sample auto-covariance can be expressed as (5). This complicated expression consists of the first right-hand-side term representative of the true auto-covariance value and three terms caused by random noise. Because of the slowly varying characteristic of the phasor, the product of the phasor and its conjugate can be neglected. Thus, the expectation of the first term is $r_1$. Also, due to the zero-mean property of the white noise, the second and third right-hand-side terms have zero expectations. In regard to the last term, since $n(m,t)$ are white noises, its expected value is zero. Hence, $C$ is an unbiased estimate, and (5) can be simplified as

$$\gamma_1 = E(C) = r_1 = kv^2, \qquad (6)$$

where $\gamma_1$ denotes the first order auto-covariance of the time series. Similarly, by denoting the zeroth order auto-covariance as $\gamma_0$, the unbiasedness of $I$ can be readily derived:

$$\gamma_0 = E(I) = v^2 + s^2, \qquad (7)$$

And the first order auto-correlation coefficient $\rho_1$ of the time series is

$$\rho_1 = \frac{\gamma_1}{\gamma_0} = \frac{kv^2}{v^2 + s^2}. \qquad (8)$$

Finally, the convergence and asymptotic distributions of sample auto-covariance and auto-correlation have been proved in time series theory [31]–[33]. When $T, M \to \infty$, the asymptotic relation between decorrelation $D$ and intensity $I$ can be given as:

$$D = 1 - \frac{C}{I} \to 1 - \rho_1 = k \cdot iSNR + 1 - k, a.s., \qquad (9)$$

where $iSNR = \frac{s^2}{I}$ is the local iSNR and a.s. denotes convergence with probability one. Particularly, for static and noise regions, (9) can be simplified as

$$D \to iSNR, a.s.. \qquad (10)$$

That indicates the decorrelation of voxel in static and noise regions is asymptotically equal to its iSNR. Furthermore, the sample estimators of both covariance and correlation converge to their real value in distribution at a moderate speed [31]–[33].

$$\sqrt{N}(I - \gamma_0) \xrightarrow{d} \xi_0, \quad as\ N \to \infty,$$
$$\sqrt{N}(C - \gamma_1) \xrightarrow{d} \xi_1, \quad as\ N \to \infty, \qquad (11)$$
$$\sqrt{N}(D - 1 + \rho_1) \xrightarrow{d} R_1, \quad as\ N \to \infty,$$

where $N = T \cdot M$ is the spatio-temporal kernel size, i.e., the total independent voxel samples used for averaging, and $\xrightarrow{d}$ means convergence in distribution. $\xi_0, \xi_1, R_1$ are zero-mean-value normal random variables. Their variances can be derived from time series theories if more assumptions are added to this model. But the expression is too complicated to be explicated here, nor will it be plausible to calculate in practice. However, (11) provides a significant support for the convergence speed of the sample auto-correlation with respect to the spatio-temporal kernel size.

The first-order sample auto-covariance is generally a complex number owing to its complex variance. And since we use real values in practice, the modulus of sample auto-covariance introduces deviation from the expected one. This problem can be addressed by modeling $|C|$ with a Rician distributed random variable [34]. $|\gamma_1|$ is its potential mean and the last three terms in (5) affect as a zero-mean-value complex normal random variable owing to the large number theorem, with $\sigma_c$ denoting its standard variance. Thus, the expectation of $|C|$ can be expressed as

$$E(|C|) = \sigma_c \sqrt{\frac{\pi}{2}} L_{\frac{1}{2}}\left(-\frac{|\gamma_1|^2}{2\sigma_c^2}\right). \qquad (12)$$

Here $L_{\frac{1}{2}}()$ is the Laguerre polynomial with order $\frac{1}{2}$. When the system SNR is sufficiently high and $|\gamma_1| \gg \sigma_c$, it is reasonable to apply approximation to the Laguerre polynomial: $\sigma_c \sqrt{\frac{\pi}{2}} L_{\frac{1}{2}}\left(-\frac{|\gamma_1|^2}{2\sigma_c^2}\right) \approx |\gamma_1|$. Then, we obtain

$$E(|C|) = |\gamma_1|. \qquad (13)$$

Numerical stimulation was performed to validate the derived ID asymptotic relation. We presumed the amplitude of static and dynamic voxel complies to Rayleigh distribution and the phase of dynamic voxel is uniformly distributed between $[-\pi, \pi]$. And the white noise is a complex Gaussian random variable [35], though these extra assumptions are not essential to this model. Moreover, SNR and the noise level are approximately equal to real ones in our experiment system. Figure 1(a) shows a schematic of OCT data with each vector denoting a complex OCT signal $X$. In Fig. 1(b), both intensity and decorrelation were calculated with a total kernel size of $N = 3960$, according to which all voxels were projected into the ID space. The asymptotic lines are plotted in the ID space using (9) as black solid lines in Fig. 1(b)-(e) with first-order correlation coefficients $k = 0, 0.5, 1$. The scatter plots report the distribution of simulated voxels. The static (and noise, blue points in Fig. 1(b)) and dynamic (red $k = 0$ and green $k = 0.5$) voxels are distributed around their asymptotic lines respectively, which well validates the asymptotic linear ID relation in (9) and (10).

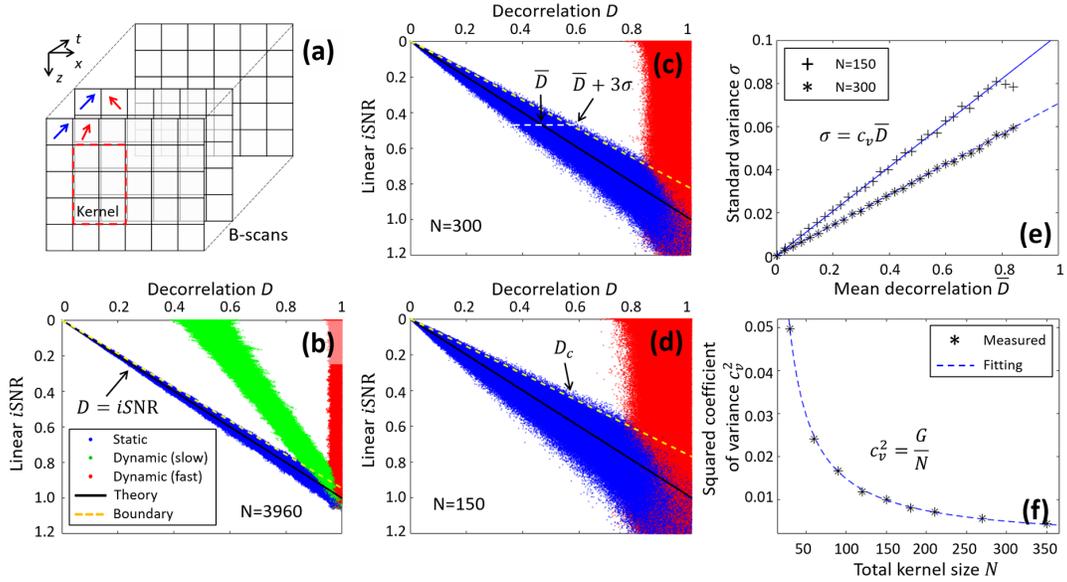

Fig. 1. Inverse SNR-decorrelation (ID) asymptotic relation and its variance. (a) Schematic diagram of OCT data. (b-d) The ID asymptotic relation. The distributions of simulated OCT voxels are marked with blue (totally static, $k = 1$), green (partially dynamic, $k = 0.5$) and red (totally dynamic, $k = 0$) points, where $k$ is the first-order correlation coefficient. The black solid lines are the theoretical asymptotic ID relation in (8), and the yellow dashed lines are the classification lines in (16). The parameters used for averaging are (b) $M = 99, T = 40, N = 3960$, (c) $M = 15, T = 20, N = 300$, (d) $M = 15, T = 10, N = 150$. $M$ is the spatial kernel size, $T$ is the number of B-scan repetitions, and $N = T \cdot M$ is the spatio-temporal kernel size. (e) The plot of standard variance $\sigma$ to mean value of decorrelation $\overline{D}$ for static voxels. Points are from numerical simulation with $N = 150, 300$ respectively and straight lines are the corresponding fitting lines with coefficients of variance $c_v \approx 0.103, 0.0707$. (f) The plot of squared coefficient of variance $c_v^2$ to spatio-temporal average kernel size $N$ for all static voxels. Points are from numerical calculation and the dashed line is the fitting result with the parameter $G \approx 1.5$.

In brief, we established a MVTS model describing the time series of OCT tomograms and derived a concise ID asymptotic relation even based on trivial assumptions excluding specific distributions. Particularly, in static and noise regions, the decorrelation is asymptotically equal to the local iSNR.

*B. Variance of asymptotic distribution*

The ID relation obtained above holds when the kernel size N is infinitely large. But in practice the voxel number used for calculation is limited, resulting in a variance of the distribution. According to (10) and (11), the variance in static and noise regions should be determined only by SNR and the spatio-temporal kernel size. Though its analytical expression is too complicated to derive here, in this section, we explored the variance numerically.

To explore the statistical properties of the variance, similar simulation was performed with a series of spatio-temporal kernels. Equation (11) predicts an asymptotic normal distribution of static and noise voxels. Basically, the standard variance, which reflects the deviation of data points from its central line, is obviously increasing when decreasing the spatio-temporal kernel size by comparing Fig. 1(b)-(d). Additionally, Fig. 1(e) illustrates the plot of standard variance of decorrelation $\sigma$ with respect to its mean value $\overline{D}$, where points with decorrelation approaching 1 are absent because the upper limit of decorrelation interferes the calculation of variance. Obviously, the standard variance $\sigma$ is almost proportional to the mean decorrelation $\overline{D}$. And a linear fitting was performed ($R^2 = 0.99$), i.e.,

$$\sigma = c_v \overline{D}, \quad (14)$$

where $c_v$ is the coefficient of variance (CoV) of decorrelation. Varying the kernel size $N$, the linear relation exists with a different slope $c_v$. The change of squared CoV $c_v^2$ over the kernel size $N$ is plotted in Fig. 1(f). According to (11), when the kernel size $N$ increases, the squared CoV decreases gradually and can be fitted by ($R^2 = 0.99$)

$$c_v^2 = \frac{G}{N}. \quad (15)$$

It is worthy to note that owing to the homogeneity of decorrelation, the CoV parameter $G \approx 1.5$ is invariant with regard to the system noise level. Thus, the variance of decorrelation in static and noise regions is merely determined by the local SNR and the spatio-temporal kernel size, and thus

can be computed given $N$ and SNR as

$$\sigma^2 = \frac{G}{N} \cdot \text{iSNR}^2. \quad (16)$$

A range of $3\sigma$ based on Pauta Criterion can be used as a boundary line as indicated by the yellow dashed lines in Fig. 1(b)-(d), i.e.,

$$D_c = \text{E}(D) + 3\sigma = \left(1 + 3\sqrt{\frac{G}{N}}\right)\text{iSNR}. \quad (17)$$

Accordingly, given a limited kernel with size $N$ in practice, the population distribution of static and noise voxels can be readily determined: a central line determined by (10) with its variance calculated by (16).

*C. ID-OCTA algorithm*

Motion contrast OCTA creates angiograms by identifying the dynamic flow and removing the static background tissue according to the temporal changes between two successive tomograms taken at the same location. Referring to Fig. 1(b), it is impossible to identify the dynamic flow simply based on a decorrelation threshold, because the low-SNR static voxel also presents a high decorrelation value, resulting in the pseudo-dynamic artifact. Conventionally, an empirical and SNR-invariant intensity threshold is used to remove noise, but meanwhile it excludes dynamic regions with low SNR, e.g., vessels in the deep. Thus, precise dynamic-static classification is desired for all SNR.

In real sample data, though vessels vary in $k$ values and correspondingly locate around different lines because of different blood speeds, the distribution of static and noise voxels is determined and invariant. Accordingly, the static and noise voxels can be totally removed at all SNR levels, and the remaining dynamic voxels are treated as blood flow and used to generate angiograms. And the boundary line established above is appropriate for classification. Besides, the $3\sigma$ can be extended depending on the extent of discrimination that users need. In practice, the noise level $s^2$ is assumed invariant along the depth and can be determined in advance by averaging the air region and the bottom noise region in tomograms. The procedure of the whole classification is summarized in Algorithm 1.

### III. EXPERIMENTAL RESULTS

In this section, ID-OCTA method was performed on both flow phantom and human skin data. The advantage of ID-OCTA is demonstrated by comparing with conventional cmOCT both quantitatively and qualitatively.

*A. System setup and implementation*

The OCTA system used in this study was mainly based on a typical spectral domain OCT, which has been detailed in [36]. Briefly, a broadband super luminescent diode (SLD) with a central wavelength of 1325 nm and a full width at half maximum bandwidth of 100 nm was used as the light source, theoretically providing an axial resolution of 7.6 μm in air. A customized lens assembly was used as an objective lens (focal length = 36 mm) to achieve a 10 μm lateral resolution. A customized high-speed spectrometer equipped with a fast line-scan InGaAs camera (120 kHz line-scan rate) was used to record the spectral interference fringes. In this work, a stepwise raster scanning protocol (z-x-y) was used for volumetric imaging over a 2.5 mm × 2.5 mm (x × y) area, with 300 A-lines per B-scan (fast-scan, x direction) and 300 B-scans repeated 5 times at each position per volume (slow-scan, y direction).

The phantom was composed of two half parts: one was solidified gel phantom mixed with low concentration intralipid to mimic static tissues; and the other was lipid solution to mimic dynamic flow. The intralipid mixture and lipid solution were prepared in a similar way to [35] with measured attenuation coefficients of ~1.6 mm$^{-1}$ and ~1.3 mm$^{-1}$ respectively. To facilitate the test of ID asymptotic relation with a large kernel, a total of 1500 B-scans were acquired at the same location. And a 4D spatio-temporal average kernel of 5 × 3 × 1 × 20 (z × x × y × t) was used to analyze the flow phantom data.

In the human skin imaging, OCT scans were performed on 6 healthy subjects with informed consents and all scans were performed on the cheek regions of the subjects. Phase compensation was applied after data acquisition by averaging phase differences between two adjacent A-lines along the depth axis [37]. As a tradeoff between the resolution and performance, a hybrid (spatial Gaussian and temporal moving average) spatio-temporal average kernel of 3 × 3 × 3 × 5 (z × x × y × t) was used in both ID-OCTA and cmOCT. In ID-OCTA, a $3\sigma$ range is used to determine classification lines. In cmOCT, the intensity threshold was set at about 2 × the noise level in order to discriminate vessels at the same depth with ID-OCTA.

To quantify the performance advantage of the proposed ID-OCTA over the conventional cmOCT, contrast-to-noise ratio (CNR) were calculated as below [28],[38]:

$$\text{CNR} = \frac{\overline{D_s} - \overline{D_n}}{\sigma_n}, \quad (18)$$

where $\overline{D_s}$, $\overline{D_n}$ refer to the mean values of decorrelation in signal and noise regions, and $\sigma_n$ denotes the standard variance in the noise region.

---

**Algorithm 1** Framework of ID feature classification

**Input:**
  The noise level of OCT system $s^2$;
  Unclassified OCT complex data $Complex(z, x, y)$;
  CoV parameter $G \approx 1.5$;
  Spatio-temporal average kernel size $N$

**Output:**
  Discriminated blood flow $Y$

1: Map data into ID space, $X = \text{ComputeID}(Complex)$
2: Establish the linear ID classifier $D_c = (1 + 3\sqrt{\frac{G}{N}})\text{iSNR}$
3: Reserve all data points below the classifier, denoting classified data as $Y$
4: Generate 3D OCTA angiograms

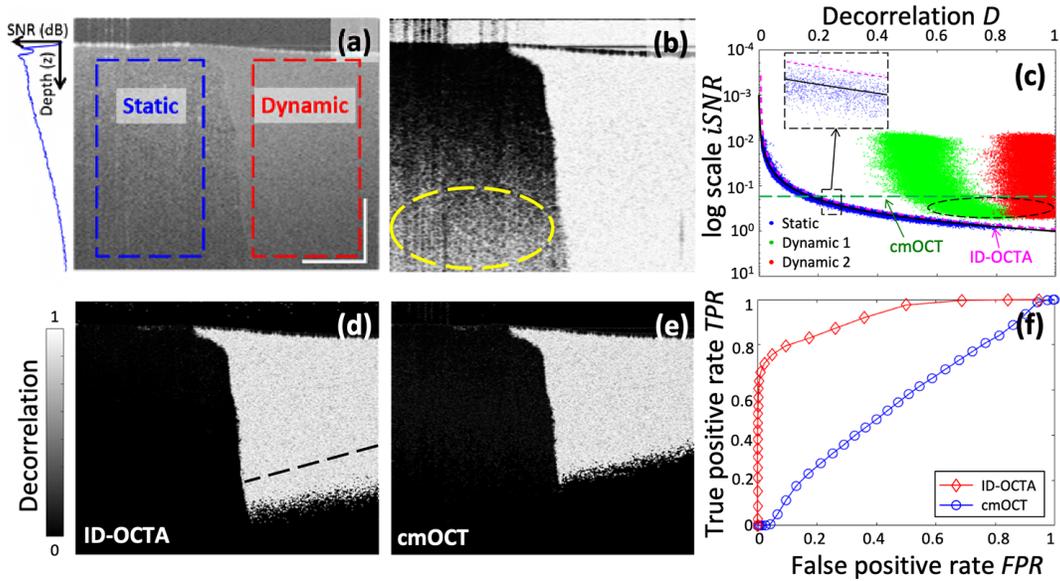

Fig. 2. Flow phantom data validate the feasibility of ID-OCTA. (a) Structural (intensity) cross section of flow phantom. Static area is the left half and flow area is the right half. The dashed boxed indicates the parts used for ID space mapping. Insert is the averaged depth profile indicating the SNR decay. (b) Decorrelation mapping of the cross section. (c) ID space mapping of the phantom data and the proposed classifier. The static and noise voxels are marked in blue and the dynamic voxels with different B-scan intervals are marked in red (9.9 ms) and in green (3.3 ms). Insert is an enlarged view of the dashed box region. The corresponding theoretical asymptotic relation in (9) is the black solid curve, the ID classifier is the magenta dashed line using (16), and the intensity threshold in cmOCT is the green dashed line. The circled area indicates flow signals excluded by cmOCT. Cross-sectional angiogram by (d) the proposed ID-OCTA, and (e) cmOCT at the same false positive rate (FPR). (f) Receiver-operating characteristic (ROC) curves of the two classifiers. TPR, true positive rate. Scale bar = 0.5 mm.

### B. Flow phantom experiment

As shown in the structural cross section in Fig. 2(a), the flow phantom offers exact prior knowledge of static (left half) and dynamic (right half) regions. To avoid the ambiguity on the static-dynamic boundary, only the rectangular regions marked with dashed boxes in Fig. 2(a) were used for the following validation, which also served as the ground truth. The decorrelation mapping was readily computed, as shown in Fig. 2(b). Generally, the dynamic region presents a high decorrelation value and the static region has a low value. However, as probing light penetrates deeper, the signal SNR decays almost exponentially, referring to the depth profile inserted in Fig. 2(a). And due to the influence of the random noise, static regions in the deep also present high decorrelation values as highlighted by the yellow ellipse in Fig. 2(b). As illustrated in Fig. 2(c), static voxels (blue points) and dynamic voxels (red and green points) with different B-scan intervals are mapped into ID space with log-scaled iSNR, where the ID relation can be perceived clearly. The static and noise voxels in blue distribute around the black curve determined by the asymptotic ID relation in (9). In view of the existence of outliers, least angle regression algorithm [39] is used to mitigate their influence in the fitting test. The curve fits quite well with phantom data ($R^2 = 0.99$). In addition, the dynamic flow presents a higher decorrelation value when acquiring successive tomograms with a larger B-scan interval, which accords with that motion sensitivity is determined by time interval.

As illustrated in Fig. 2(c), the proposed ID classifier (magenta dash line) was determined by (16) and a uniform intensity threshold (green dashed line) was set for the conventional cmOCT to remove low-SNR regions. It is obvious that cmOCT excludes flow signals in the low SNR region highlighted by the black circle in Fig. 2(c), which is preserved by ID-OCTA. Comparing the two cross-sectional angiograms of Fig. 2(d) and 2(e), ID-OCTA apparently presents a higher visibility in the deep dynamic area (below the black dashed line) than cmOCT.

To evaluate the classification performance of ID-OCTA quantitatively, we compute the true positive rate (TPR) and false positive rate (FPR) of classification result as follows:

$$\text{TPR} = \frac{T_p}{T_p + F_p},$$

$$\text{FPR} = \frac{F_p}{F_p + T_n}, \quad (19)$$

where $T_p$, $F_p$ and $T_n$, respectively, are the number of true positives, false positives and true negatives of classification. The classification was performed on static and flow data points with the longer time interval, and the receiver-operating characteristic (ROC) curve is presented to reflect the performance of the proposed classifier and conventional SNR-invariant threshold. As Fig. 2(f) displays, the advantage of the ID classifier is clear: to obtain a TPR of about 80%, ID-OCTA has an FPR of 9.9% whereas cmOCT is 75.7%, which means ~86.9% improvement in misclassification rate. Decreasing the kernel size to 5 × 3 × 1 × 9 leads to an increased FPR, whereas ID-OCTA at 11.8% still has ~84.7% improvement in contrast to cmOCT at 77.4%. The CNR of ID-OCTA in Fig. 2(d) is about 39.6, much higher than 17.3 of cmOCT in Fig. 2(e), indicating ~2.3 times improvement in vascular contrast.

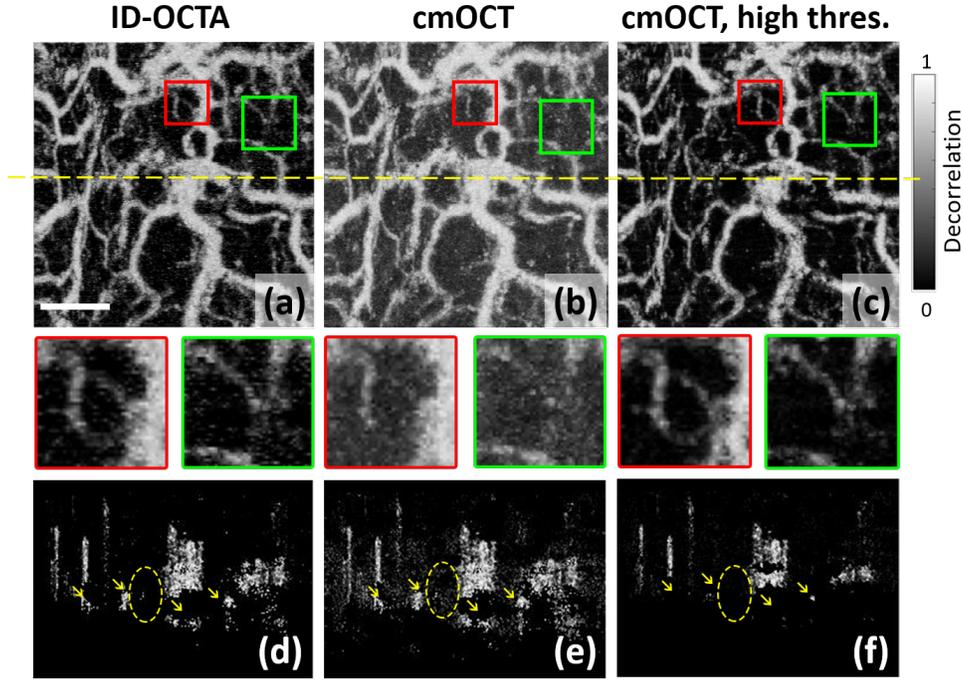

Fig. 3. ID-OCTA in vivo imaging on human skin compared with cmOCT. En-face angiograms generated by (a) ID-OCTA, (b) cmOCT and (c) cmOCT with a high intensity threshold at about 6 × the noise level. Inserts are enlarged views of the enclosed areas respectively. (d)-(f) Corresponding cross-sectional angiograms along the yellow dash line. Scale bar = 0.5 mm.

*C. In vivo human skin imaging*

As shown in Fig. 3(a), ID-OCTA enables a clear visualization of the blood perfusion within the high scattering skin tissue. Compared with ID-OCTA (Fig. 3(a)), cmOCT (Fig. 3(b)) exhibits higher background noise, which is mainly induced by the residual static voxels in the low-SNR areas as circled by yellow ellipses in Fig. 3(d) and 3(e). And this noisy background impairs the visibility of small blood vessels as shown by rectangles in Fig. 3(a) and 3(b) and the enlarged views. On the other hand, as displayed in Fig. 3(c), though elevating the intensity threshold suppresses the background noise such that the contrast is comparable to ID-OCTA, it also excludes blood flow signals in low-SNR regions as pointed out by arrows in Fig. 3(f), degrading the vascular visibility in the deep. In agreement with the flow phantom experiment, ID-OCTA allows precise discrimination of dynamic flow from static background tissues at all SNR level compared to cmOCT, particularly, an enhanced visualization of blood vessels in the deep.

To further quantify the contrast improvement of ID-OCTA over cmOCT, noise and vessel areas are labelled by experienced dermatologists as shown in Fig. 4(a). In Fig. 4(b) and (c), the pseudo-color en-face angiogram of cmOCT is apparently noisier than ID-OCTA. Fig. 4(d) and (e) further validate the contrast improvement by plotting the histograms of angiograms generated by two methods, where ID-OCTA obviously separates vessels and background better. Most likely due to the noisy background in cmOCT, the center of the noise histogram has an obvious bias from zero, and the low-flow component of the vessel histogram has been pushed up, resulting in a narrowed dynamic range in cmOCT. According to (18), the CNR of ID-OCTA (5.96 ± 1.00) is significantly higher than that of cmOCT (5.03 ± 0.89) based on the paired sample t-test on 6 samples ($P-\text{value} = 5.75 \times 10^{-4}$), corresponding to ~18.5% improvement in vascular contrast.

In addition, the proposed ID-OCTA also enables an improved accuracy of vessel density quantification. The vessel density was quantified by applying a decorrelation threshold to both ID-OCTA and cmOCT en-face angiograms [29], and the pixels above the threshold divided by the total pixels in the region of interest was the vessel density. Using the manually labelled mask as the ground truth for vessels, the relative error (RE, relative difference between the measured vessel density and the ground truth) of the vessel density was calculated. The RE of ID-OCTA (26.6% ± 0.3%) is significantly lower than that of cmOCT (31.9% ± 0.4%), corresponding to ~19.6% improvement in the quantification of vessel density.

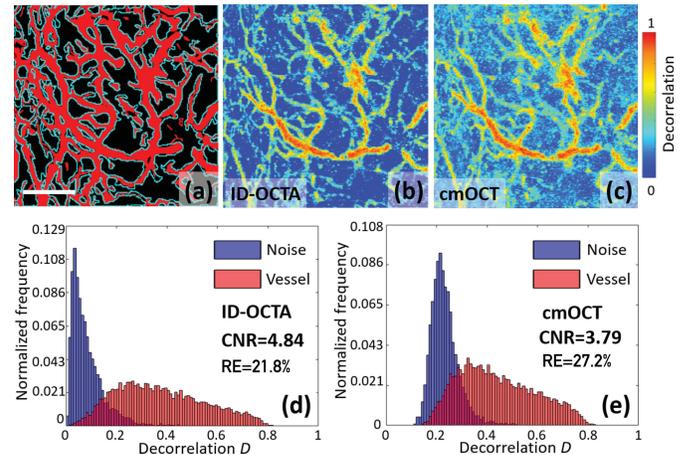

Fig. 4. Qualitative comparison of ID-OCTA to cmOCT on sample 4. (a) Vessel and noise area masks. The red area refers to vessels while noise areas are enclosed by cyan contours. En-face projections of (b) ID-OCTA and (c) cmOCT. Histograms of vascular signal and background noise in (d) ID-OCTA and (e) cmOCT. Scale bar = 0.5 mm.

## IV. DISCUSSION

Decorrelation-based OCTA has been widely used in clinical diagnosis and assessment, but the decorrelation-SNR dependence severely degrades the vascular visibility, particularly in the low-SNR regions. To solve this problem, we proposed a novel, SNR-adaptive ID-OCTA method enabled by the statistical properties of the relation between iSNR and decorrelation. The statistical characteristics of the relation is revealed by modeling OCT tomograms with MVTS theory. Derived from the MVTS model, the decorrelation has a universal asymptotic linear relation to iSNR for both static and dynamic signals. Particularly, the variance of static and noise voxels is solely determined by the spatio-temporal kernel size at a given SNR level. Thus, the population distribution of static voxels can be estimated in ID space given the spatio-temporal average kernel size and then be removed to generate precise classification of blood vessels from surrounding tissues, which is basically an analogy to the chemical erosion of surrounding tissues in vascular casting. Then numerical simulation and flow phantom experiment were performed to validate the aforementioned conclusions. The proposed ID-OCTA was further compared with the conventional cmOCT using human skin data, which shows the proposed ID-OCTA is able to reveal vasculature in the deep and offer better contrast of angiograms compared with cmOCT.

First of all, the propose ID-OCTA method offers several advantages in vascular contrast enhancement. Rather than using modified decorrelation algorithms with SNR correction [27] or noise-bias correction [20], the proposed ID-OCTA directly uses the decorrelation without complicated noise correction, and solves the SNR-dependence of decorrelation by the thorough understanding of this dependence, including the ID asymptotic relation and variance. ID-OCTA handles not only the noise-induced bias of decorrelation but also the variance of decorrelation due to the limited kernel size. In addition, only the system noise level and the spatio-temporal kernel size are required in ID-OCTA without additional measurement of system parameters nor complicated estimation. Different from the numerical approach by Gao et al.[29], we comprehensively consider the influence of factors like intensity, system noise and the average kernel size. Thus, ID-OCTA enables a thorough understanding of the statistical features of intensity and decorrelation, as well as better generality and adaptability. Different from the learning-based method [16], the classification map is directly determined using the statistical properties of the ID relation in ID-OCTA, which avoids the influence of differences between flow phantom and real blood flow and allows a fast switch between different parameter settings without repeated training. In addition, on a dataset with 512 × 300 × 300 voxels and 5 B-scan repetitions, ID-OCTA classification took less than 1 s for generating a final OCT angiogram, which was tested on a MacBook Pro with an Intel i5 CPU at 2.30 GHz and 16 GB memory. Although an additional time-consuming phase compensation (~15 s) is currently required due to the use of 4D kernel, the clinical translation of the proposed ID-OCTA can be achieved by employing GPU to accelerate image processing (e.g. OMAG).

Secondly, ID-OCTA implies a desire for higher system SNR level because high SNR could diminish the variance of the asymptotic distribution. Besides, though the linear classifier in ID space is theoretically invariant with respect to SNR, the Rician model aforementioned requires a high SNR for unbiased estimation. In OCT data, the SNR decay is mainly caused by the light attenuation over the penetration depth, the sensitivity rolling-off in Fourier domain OCT, and the limited depth of field. Accordingly, optical clearing agent (e.g. glycerol solutions) can be used to enhance the transparency in skin. Swept-source OCT usually has an improved spectral resolution and offers a superior performance in sensitivity rolling-off. And the extended focus design [40] and dynamic focusing [41] would also be helpful to alleviate the influence of defocus.

Thirdly, the ID relation could further guide the optimal design of imaging system and signal processing algorithms. A convergent distribution of the static voxels, i.e., a small variance $\sigma^2$ in the ID space is desired in order to alleviate the overlap between the static and dynamic voxel population, e.g., Fig. 1(d) has more overlap areas than Fig. 1(b). According to (14), the benefit of increasing averaged voxel number is self-evident for accelerating the convergence of correlation coefficient. Nonetheless, an enlarging kernel size would degrade the spatial and temporal resolution of OCT angiogram. For another, a large number of repetitions may result in a long imaging time and severe bulk motion, which further lead to unexpected phasor changes and violate the slowly-variant assumption of the phasor. Thus, a trade-off must be taken into consideration when selecting the kernel parameters. An effective approach might be collecting voxel samples in additional dimensions. In our 3D skin imaging, a 4D spatio-temporal kernel of 3 × 3 × 3 × 5 was applied. Its degradation in the spatio-temporal resolution is equivalent to a spatial 3 × 3 × 3 Gaussian smoothing kernel and 5 repeated B-scan in the conventional OCTA methods, which indicates no excess degradation of spatial and temporal resolution in our ID-OCTA. The collection of voxel samples can be further extended to the wavelength (e.g. splitting spectrum [10], [42]), polarization [20], [27] and even the angular (e.g. splitting full-space B-scan modulation spectrum [43]) dimensions. In addition, a multi-dimensional kernel would also offer more degrees of freedom to balance the OCTA performance for a given total kernel size.

Finally, in addition to improving the visibility of vascular morphology, the statistical properties of ID features are useful for the hemodynamic quantification. Though the decorrelation index has been used for quantitative hemodynamic analysis [21], [25], it is highly related to the local SNR according to (8), i.e., low-SNR blood flow presents a higher decorrelation value. Thus, the asymptotic ID relation indicates a necessity of SNR-correction for decorrelation, otherwise it will pose a misinterpretation of the outcomes. Besides, it is suggested that the time interval between adjacent tomograms determines the motion sensitivity and the dynamic range of the motion measurement.

## V. CONCLUSION

To solve the dependence of decorrelation on SNR, we develop a MVTS model to provide a mathematical understanding of their relation. Based on the model, we found that decorrelation has an asymptotic linear relation to iSNR with a variance determined by the spatio-temporal kernel size *N* in static tissues. By calculating the variance numerically and removing static and noise voxels correspondingly, a SNR-adaptive classifier using ID features is established. Validated on the flow phantom, the proposed ID-OCTA presents higher classification accuracy, especially for low-SNR regions. Based on human skin experiments and the comparison with cmOCT, we further corroborate its advantages in visualizing vessels in the deep and enhancing the visibility as well as the contrast of angiograms. And implications of this work on both system design and hemodynamic quantification are discussed. The proposed ID-OCTA can hopefully benefit the diagnosis and assessment of diseases and expand the applicability of OCTA in clinic.